\let\oldyear\year
\documentclass{IEEEtran}

\let\year\oldyear
\usepackage{cite}
\usepackage{amsmath,amssymb,amsfonts}
\usepackage{graphicx, pstricks}
\usepackage{comment}

\usepackage{textcomp}

\usepackage{algorithm}
\usepackage{algpseudocode}
\usepackage{url}

\usepackage{multirow}
\newcommand{\shorten}[1]{}

\newtheorem{theorem}{Theorem}
\newtheorem{definition}{Definition}

\newtheorem{lemma}{Lemma}

\newtheorem{corollary}{Corollary}
\newtheorem{example}{Example}

\def\BibTeX{{\rm B\kern-.05em{\sc i\kern-.025em b}\kern-.08em
    T\kern-.1667em\lower.7ex\hbox{E}\kern-.125emX}}

\begin{document}

%\title{Modeling Network Slicing in 5G by Combinatorial Designs}
%      or 
\title{Expanded Combinatorial Designs as Tool to Model Network Slicing in 5G}

	\author{
	\IEEEauthorblockN{Danilo Gligoroski and Katina Kralevska\\
		Email:  \{danilog, katinak\}@ntnu.no}\\
	\IEEEauthorblockA{\IEEEauthorrefmark{1}Department of Information Security and Communication Technologies, Norwegian University of Science and Technology, 
		Trondheim, Norway}
	}

	\maketitle

\begin{abstract}
The network slice management function (NSMF) in 5G has a task to configure the network slice instances and to combine network slice subnet instances from the new-generation radio access network and the core network into an end-to-end network slice instance. In this paper, we propose a mathematical model for network slicing based on combinatorial designs such as Latin squares and rectangles and their conjugate forms. We extend those designs with attributes that offer different levels of abstraction. For one set of attributes we prove a stability Lemma for the necessary conditions to reach a stationary ergodic stage. We also introduce a definition of utilization ratio function and offer an algorithm for its maximization. Moreover, we provide algorithms that simulate the work of NSMF with randomized or optimized strategies, and we report the results of our implementation, experiments and simulations for one set of attributes.
\end{abstract}

{\bfseries {Keywords}}:
5G networks, Combinatorial designs, Dynamic deployment, Latin squares, Latin rectangles, Network slicing, Optimal slice selection.

\IEEEpeerreviewmaketitle

\section{Introduction}
\label{sec:introduction}
In the global market capitalization, 5G technologies are projected to be worth over USD 12.3 trillion by 2035 \cite{kim2018network}, and network slicing is seen as the key enabling technology that can bring up to 150\% increased revenues for the operators, in comparison with the classical one-big network concept \cite{EricssonAndBT-17}. The idea for network slicing in 5G came from telecommunication industry alliance NGMN in February 2015 \cite{alliance20155g} and very shortly afterwards was accepted by 3GPP \cite{3gpp2015-22-891} as an enabling technology that will bring new services and markets. 

\begin{figure}
	\includegraphics[width=3.3in]{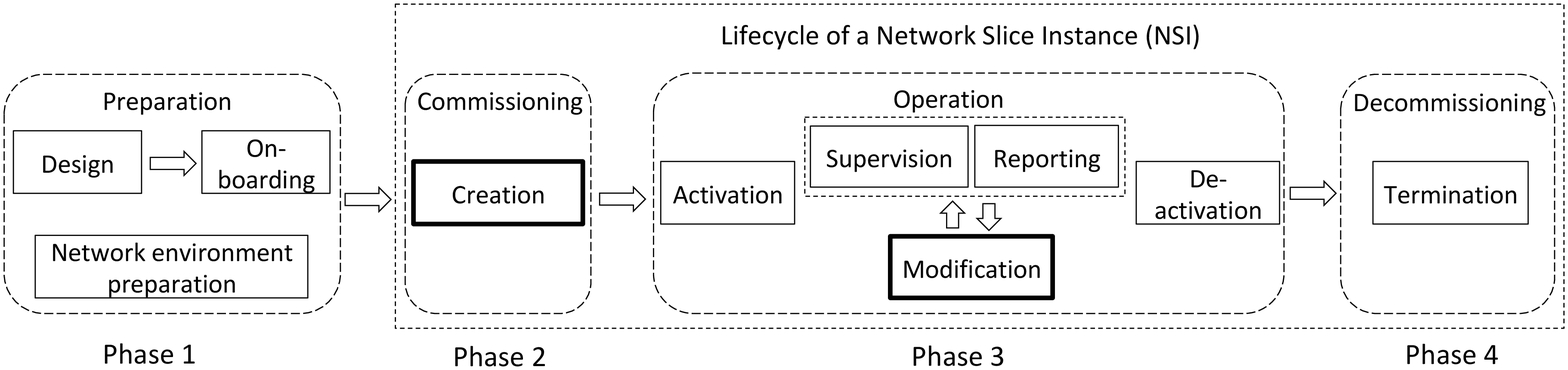}
	\caption{Management aspects of network slice instance \cite{3gpp2}. We propose a model based on combinatorial designs for the creation and modification steps (represented with thick frames).}
	\label{lifecycle}
\end{figure}

The role of network slicing is to enable functional and operational diversity on a common network infrastructure \cite{3gpp}. %\cite{7926920,3gpp}
The idea is to create multiple isolated networks, termed Network Slice Instances (NSIs), on a common physical infrastructure where physical and virtual resources of each NSI are customized to satisfy the requirements for a specific communication service. 
Fig. \ref{lifecycle} presents the management phases of a NSI: 1. preparation; 2. commissioning; 3. operation; and 4. decommissioning. 
The preparation phase includes all steps required before the creation of a NSI (creation and verification of network slice template, evaluation of network slice requirements, capacity planning). The lifecycle of a NSI starts with the second phase. 
During the commissioning phase, the NSI is created and all resources for the NSI are allocated and instantiated.
In the operation phase, the NSI supports a communication service. First, the NSI is activated and later performance reporting for KPI monitoring as well as modification and de-activation of the NSI happen. 
The last phase of NSI lifecycle and NSI management includes termination of the NSI by releasing the dedicated resources and removing the NSI specific configuration from the shared resources. After this phase, the NSI does not exist anymore.

\begin{figure}
	\includegraphics[width=3.3in]{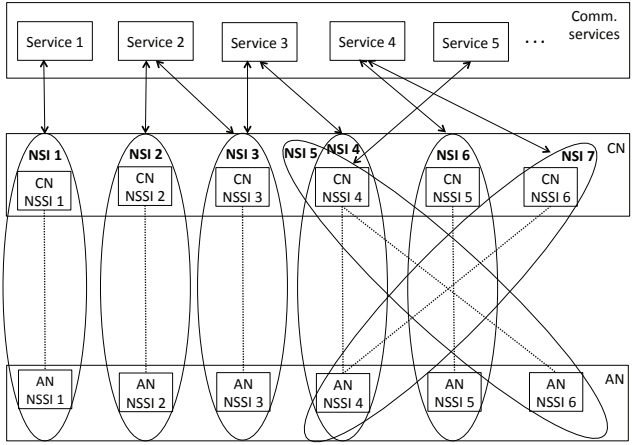}
	\caption{Five services supported by seven NSIs. The NSIs contain NFs, belonging to CN and AN NSSIs, and the interconnection information between the NFs.}
	\label{architecture}
\end{figure}

The slicing is performed end-to-end (E2E) \cite{DBLP:journals/corr/LiWPU16,8334921}. Thus, a NSI contains Network Slice Subnet Instances (NSSIs) in the New-Generation Radio Access Network (AN) and the Core Network (CN), refereed to as AN and CN NSSIs in Fig. \ref{architecture}, and the interconnections between them. NSSI is a set of network functions (NFs) which can be physical NFs or virtualized NFs. If the NFs are interconnected, the 3GPP management system contains the information relevant to the connections between these NFs such as topology of connections and individual link requirements. Fig. \ref{architecture} shows that one NSI may support a single (e.g. NSI 1) or multiple communication services (e.g. NSI 3). AN and CN NSSIs can be dedicated to one NSI (e.g. CN NSSI 1) or shared by two or more NSIs (e.g. CN NSSI 4). %Different NSIs contain NFs, belonging to NSSIs, as well as the information relevant to the interconnection between NFs.

A demanding tenant issues a communication service request which is translated into a slice request (network functions and infrastructure requirements) for the Mobile Network Operator (MNO). The following management functions manage the NSIs to support communication services: Communication Service Management Function (CSMF), Network Slice Management Function (NSMF) and Network Slice Subnet Management Function (NSSMF).
CSMF receives the communication service related requirements by the tenant and converts them into network slice related requirements which are sent to NSMF. NSMF manages and orchestrates the NSI. It configures the NSIs and knows which NSSIs are associated with each NSI (cross-domain management and orchestration (M\&O)). One NSSI can be associated with multiple NSIs where NSSMF manages and orchestrates the NSSIs.
The network slice is instantiated and configured by NSMF where NSMF manages the interactions among the slice instances in terms of resources and features sharing (cross-slice M\&O).
For instance, in Fig. \ref{architecture}, both AN NSSI 1 in the access part and CN NSSI 1 in the core part first have to be defined and instantiated. Then NS 1 is instantiated by combining these two NSSIs.

In spite of the vast number of articles devoted to network slicing, it comes as a surprise that there are still no general precise mathematical models for network slicing and building such models is a challenging task as suggested in \cite{ABIResearchAndIntel2018,kim2018network}. 
Moreover, even the taxonomy used by different standardization organizations (for example 3GPP and IETF) is not agreed, although they are addressing the same slicing scenarios. For example what is referred as "hard slicing" by IETF, is referred as non-shared network slice subnet instance by 3GPP (see Definition \ref{def:HardNetworkSlicingIETF} and Definition \ref{def:Non-shared3GPP} below). Similarly, "soft slicing" by IETF (Definition \ref{def:SoftNetworkSlicingIETF}) corresponds to "shared constituent of network slice instance" (Definition \ref{def:Shared3GPP}) by 3GPP.

\begin{definition} [IETF \cite{geng2017network}]\label{def:HardNetworkSlicingIETF}
\emph{Hard slicing} refers to the provision of resources in such a way that they are dedicated to a specific network slice instance.
\end{definition}
\begin{definition} [3GPP \cite{3gpp1}]\label{def:Non-shared3GPP}
A NSSI that is dedicated to one NSI and is not shared as a constituent by two or more NSSI(s) is called \emph{a non-shared NSSI}. %A NSSI may contain core network (CN) functions only or access network (AN) functions only or both CN functions and AN functions.
\end{definition}

\begin{definition} [IETF  \cite{geng2017network}]\label{def:SoftNetworkSlicingIETF}
\emph{Soft slicing} refers to the provision of resources in such a way that whilst the slices are separated such that they cannot statically interfere with each other, they can interact dynamically, which means they may compete for some particular resource at some specific time. 
\end{definition}

\begin{definition} [3GPP \cite{3gpp1}]\label{def:Shared3GPP}
A NSSI may be shared by two or more NSIs, this is called \emph{a shared constituent of NSI}. A NF may be shared by two or more NSSI(s), in which case it is called \emph{a shared constituent of NSSI}. %A NSSI may be shared by two or more NSIs, this is also called a shared constituent of NSI.
\end{definition}

\subsection{Related Work}

The ideas for network slicing originates from the areas of Cloud Computing \cite{regalado2011coined}, Software Defined Networks (SDN) proposed by IETF \cite{yang2004forwarding}, Network Functions Virtualisation (NFV) \cite{etsi2013001} and Information-Centric Networking (ICN) \cite{ghodsi2011information}. One of the major research problems is the resource allocation across slices. Several works address the slicing of radio access network resources or cross-domain on VNF level. We mention here some of the most prominent mathematical models developed for network slicing. 

Reference \cite{8329496} presents a mathematical model to construct network slice requests and to map them on the network infrastructure. The mapping process is performed on VNF level where first it places the VNFs to the nodes in the network and later it selects that paths between the VNFs and chains them.
With the aim to maximize the long-term network utility, reference \cite{8382171} uses a genetic algorithm to serve slice requests. 

Network slicing brings new business models and interactions between the infrastructure providers, the tenants and the customers. This opens many directions for optimizations.
The algorithm for admission and allocation of network slices requests in \cite{8057045} maximizes the infrastructure provider's revenue and ensures that the service guarantees provided to tenants are satisfied.

\subsection{Our Contribution}
In this paper, we offer one mathematical model for the Network Slice Management Function (NSMF) based on combinatorial designs and their algebraic properties. 
We see our contribution as one step closer to a general, precise and scalable mathematical model for network slicing. In particular, our mathematical model addresses the tasks of the NSMF in the creation and modification sub-phases of the NSI lifecycle (phases 2 and 3 in Fig. \ref{lifecycle}). 
The model uses combinatorial objects known as Latin squares (or Latin rectangles) to describe communication services and the NSSIs. Combinatorial designs \cite{Colbourn:2006:HCD:1202540} have been used for a long time in communications, networking and cryptography. References \cite{10.1007/978-3-642-40552-5_15,6875415,7034478} apply combinational designs for network coding. The authors in \cite{Colbourn1999ApplicationsOC} listed thirteen application areas of combinatorial designs, and in this paper we extend the list with one more application, i.e., configuration of network slices in 5G. % \cite{COLBOURN1992193}, \cite{conf/dimacs/EricksonC94}. 
The mathematical properties of our model guarantee conflict resolution for services defined over network slices that compete for resources in CN and AN, as long as the configuration and modification of NSI and NSSI are performed within our model. 

The next contribution of this paper is from an optimization point of view. We introduce the notion of utilization ratio function, with aims to describe the functional dependencies between the number of used network resources and the waiting time for establishing the network slice. We present two strategies for the work of NSMF, a non-optimized first-come-first-serve strategy and an optimal strategy, where the optimization objectives are: 1. to maximize the utilization of the network components; and 2. to decrease the average delay time from slice request to slice activation. 

Finally, we show some simulation results. The optimal strategy achieved by maximizing the utilization ratio function, provides more than twice better performance in terms of the both objectives compared to the non-optimized strategy.

%Although network slicing holds major promise in supporting a wide variety of use cases in 5G networks, a lot of questions still have to be addressed to enable it.
%\subsection{Our Contribution}
The rest of this paper is organized as follows. In Section \ref{example}, we give examples of modeling network slicing with combinatorial designs. In Section \ref{sysModel}, we develop general and extended combinatorial designs model for cross-domain end-to-end network slicing that includes both hard and soft slicing. In Section \ref{simul}, we instantiate our general model with several concrete attributes and present algorithms for simulation and optimization of a NSMF for that model. Section \ref{conc} concludes the paper.

\section{Examples of cross-domain network slices} \label{example}
Fig. \ref{architecture} shows five services that are provided on the same infrastructure. The resources in the access network part, such as bandwidth, computing and storage, are represented with 6 AN NSSIs, whereas the resources in the core network part are represented with 6 CN NSSIs. AN and CN NSSIs can be associated with one or multiple NSI(s). %The NSMF is responsible for configuring the NSIs. 

\begin{table}[]
\caption{A rectangular scheme, with services as rows, AN NSSIs as columns, and CN NSSIs as table entries, representing the E2E slicing described in Fig. \ref{architecture}.}
    \centering
\begin{tabular}{|c|l||c|c|c|c|c|c|}
\hline
 & & \multicolumn{6}{c|}{AN NSSIs} \\ \cline{2-8}
\hline
 & & $a_1$ & $a_2$ & $a_3$ & $a_4$ & $a_5$ & $a_6$ \\
\hline  \hline
S       & $s_1$  & $c_1$ &       &       &       &       &       \\ \cline{2-8}
E       & $s_2$  &       & $c_2$ & $c_3$ &       &       &       \\ \cline{2-8}
R       & $s_3$  &       &       & $c_3$ & $c_4$ &       &       \\ \cline{2-8}
V       & $s_4$  &       &       &       & $c_6$ & $c_5$ &       \\ \cline{2-8}
.       & $s_5$  &       &       &       &       &       & $c_4$ \\ \cline{2-8}
\hline
\end{tabular}
\label{table:UxA}
\end{table}

Let us denote the set of 5 services by $S = \{ s_1, \ldots, s_5 \} $, the set of 6 AN NSSIs by $A = \{ a_1, \ldots, a_6 \}$ and the set of 6 CN NSSIs by $C = \{ c_1, \ldots, c_6 \}$.
For this concrete example, we can represent the service/NSI/NSSI mapping as a $5 \times 6$ rectangular scheme given in Table \ref{table:UxA}. The services are modeled as rows, and the columns represent the network subnet slices of the access network part. %Thus, we have a Latin square with 5 rows where the $i-$th row corresponds to the $i-$th user and 6 columns where the $j-$th column corresponds to the $j-$th access slice. 

We fill in the rectangular scheme with elements from the set $C$. For instance, AN NSSI 6 with CN NSSI 4 forms an end-to-end slice (NSI 5) for service 5. We model this in the rectangular scheme by putting $c_4$ in the row $s_5$ and the column $a_6$. For service 4 there are two scheduled subnet slices in the access network: $a_4$ is combined with the $6-$th core network subnet slice $c_6$ and $a_5$ that is combined with the $5-$th core network slice $c_5$. We model this by placing $c_6$ in row $s_4$ and column $a_4$, and by placing $c_5$ in row $s_4$ and column $a_5$. 

Note that this configuration is for time slot $t$. The mapping scheme might change at time slot $t+\Delta t$.

When we apply dedicated resource allocation, then neither the same AN NSSI nor CN NSSI can be scheduled for more than one NSI, i.e., one service. In terms of the rectangular scheme in Table \ref{table:UxA} that means that no $c_i$ appears more than once in any column. In other words, a bundle of dedicated resources is allocated.

On the other hand, we can see that we have two $c_3$ in the $3-$rd column $a_3$ and in rows $s_2$ and $s_3$. That means that service 2 and service 3 share the $3-$rd access slice $c_3$. This is a situation when we have shared resources, i.e., soft slicing where the users compete for the resources. 

\begin{table}[]
\caption{A rectangular scheme, with core network resources as rows, RAN slices as columns, and services as table entries, representing the slicing described in Fig. \ref{architecture}.}
    \centering
\begin{tabular}{|r|l||c|c|c|c|c|c|}
\hline
 & & \multicolumn{6}{c|}{AN NSSIs} \\ \cline{2-8}
\hline
 & & $a_1$ & $a_2$ & $a_3$ & $a_4$ & $a_5$ & $a_6$ \\
\hline  \hline
\       & $c_1$  & $s_1$ &       &            &       &       &       \\ \cline{2-8}
\       & $c_2$  &       & $s_2$ &            &       &       &       \\ \cline{2-8}
C       & $c_3$  &       &       & $s_2, s_3$ &       &       &       \\ \cline{2-8}
N       & $c_4$  &       &       &            & $s_3$ &       & $s_5$ \\ \cline{2-8}
\       & $c_5$  &       &       &            &       & $s_4$ &       \\ \cline{2-8}
\       & $c_6$  &       &       &            & $s_4$ &       &       \\ \cline{2-8}
\hline
\end{tabular}
\label{table:CxA}
\end{table}

Another way of modeling the network slicing architecture is the rows to represent the core slices, the columns to represent the access slices and services are the entries in the table, as it is presented in Table \ref{table:CxA}. In the case when we want to have exclusivity, for instance one NSI for low latency and ultra reliable service, then we allocate a specific subnet slice only to one service, i.e., the services are placed in the table exactly only once in each row and column. We will elaborate this later with one precise theorem.

\begin{table}[]
\caption{A rectangular scheme, with service as rows, core network resources slices as columns, and RAN slices as table entries, representing the slicing described in Fig. \ref{architecture}.}
    \centering
\begin{tabular}{|c|l||c|c|c|c|c|c|}
\hline
 & & \multicolumn{6}{c|}{CN} \\ \cline{2-8}
\hline
 & & $c_1$ & $c_2$ & $c_3$ & $c_4$ & $c_5$ & $c_6$ \\
\hline  \hline
S       & $s_1$  & $a_1$ &       &       &       &       &       \\ \cline{2-8}
E       & $s_2$  &       & $a_2$ & $a_3$ &       &       &       \\ \cline{2-8}
R       & $s_3$  &       &       & $a_3$ & $a_4$ &       &       \\ \cline{2-8}
V       & $s_4$  &       &       &       &       & $a_5$ & $a_4$ \\ \cline{2-8}
.       & $s_5$  &       &       &       & $a_6$ &       &       \\ \cline{2-8}
\hline
\end{tabular}
\label{table:UxC}
\end{table}

Finally, for a completeness, we present the third rectangular scheme (conjugate to the previous two), with services as rows, CN NSSIs as columns, and AN NSSIs as table entries in Table \ref{table:UxC}.

\section{Combinatorial model of network slicing} \label{sysModel}
We start with some basic definitions about Latin squares and related combinatorial structures.
\begin{definition}
\emph{A Latin square} of order $n$ is an $n \times n$ array in which each cell contains a single symbol from a $n$-set $S$, such that each symbol occurs exactly once in each row and exactly once in each column.
%A Latin square of order $k$ with entries from an $k$-set X is an $k\times k$ array $L$ in which every cell contains an element of X such that every row of $L$ is a permutation of X and every column of $L$ is a permutation of X.
\label{def:Latinsquare}
\end{definition}

\begin{definition}
A $k\times n$ \emph{Latin rectangle} is an $k\times n$ array (where $k \leq n$) in which each cell contains a single symbol from a $n$-set $S$, such that each symbol occurs exactly once in each
row and at most once in each column.
\label{deLatinrectangle}
\end{definition}

\begin{definition}
\emph{A partial Latin square (rectangle)} is a square (rectangular) array $L$ with cells that are either empty or contain exactly one symbol such that no symbol occurs more than once in any row or column.
\label{PartialLatin}
\end{definition}

\begin{figure}[!h]
\minipage{0.17\textwidth}
$  \begin{bmatrix}
 1 & 3 & 5 & 2 & 4 \\
 4 & 2 & 3 & 1 & 5 \\
 3 & 1 & 4 & 5 & 2 \\
 5 & 4 & 2 & 3 & 1 \\
 2 & 5 & 1 & 4 & 3 \\
\end{bmatrix}$
\endminipage
\minipage{0.17\textwidth}
$  \begin{bmatrix}
 1 & 3 & 5 & 2 & 4 \\
 4 & 2 & 3 & 1 & 5 \\
 3 & 1 & 4 & 5 & 2 \\
\end{bmatrix}$
\endminipage
\minipage{0.17\textwidth}%
$  \begin{bmatrix}
 1 &   &   &   & 4 \\
   & 2 & 3 &   &   \\
   &   & 4 &   &   \\
 5 & 4 &   & 3 & 1 \\
\end{bmatrix}$
\endminipage
  \caption{A $5\times 5$ Latin Square, a $3\times 5$ Latin rectangle and a partial $4 \times 5$ Latin rectangle.}
\label{fig:LatinExamples}
\end{figure}
In Fig. \ref{fig:LatinExamples} we show an example of a $5\times 5$ Latin Square, a derived $3\times 5$ Latin rectangle and a derived partial $4 \times 5$ Latin rectangle.

\begin{definition}\label{def:conjugates}
Let $L$ be a $n \times n$ Latin square on symbol set $E_3$, with rows indexed by the elements of a $n$-set $E_1$ and columns indexed by the elements of a $n$-set $E_2$. Let us define a set of triplets $T = \{(x_1, x_2, x_3) : L(x_1, x_2) = x_3\}$. Let $\{a, b, c\} = \{1, 2, 3\}$. The $(a, b, c)$-\emph{conjugate} of $L$, $L_{(a,b,c)}$, has rows indexed by $E_a$, columns by $E_b$, and symbols by $E_c$, and is defined by $L_{(a,b,c)}(x_a, x_b) = x_c$ for each $(x_1, x_2, x_3) \in T$.
\end{definition}

Instead of using some general symbol sets $E_1$, $E_2$ and $E_3$ in Definition \ref{def:conjugates}, and in the rest of this paper let us use the set of services  $E_1 \equiv S = \{ s_1, \ldots, s_{n_s} \} $, the set of AN NSSIs $E_2 \equiv A = \{ a_1, \ldots, a_{n_a} \}$ and the set of CN NSSIs $E_3 \equiv C = \{ c_1, \ldots, c_{n_c} \}$. In this context, we write $(S,A,C)-$conjugate instead of $(1,2,3)-$conjugate, $(S,C,A)-$conjugate instead of $(1,3,2)-$conjugate and $(C,A,S)-$conjugate instead of $(3,2,1)-$conjugate.

In the light of our introduced mathematical formalism that uses the combinatorial objects of Latin squares and rectangles, instead of the descriptive Definition \ref{def:HardNetworkSlicingIETF} for hard slicing and its equivalent Definition \ref{def:Non-shared3GPP} for dedicated (non-shared) slice subnet instances we offer another definition for hard network slicing in the core and access parts.

\begin{definition}[Hard Core Network Slicing]\label{def:HardCoreNetworkSlicing}
\emph{Hard network slicing of $C$} is a set of triplets $T_{hard, C} = \{(s_i, a_j, c_k) : s_i \in S, a_j \in A, c_k \in C\},$ such that for any two triplets  $(s_{i_1}, a_{j_1}, c_{k_1}), (s_{i_2}, a_{j_2}, c_{k_2}) \in T_{hard, C}$ it holds: 
\begin{equation}\label{eq:HardCoreNetworkSlicing}
\left\{
\begin{array}{rl}
\text{if } s_{i_1} = s_{i_2} & \text{then } a_{j_1} \neq a_{j_2} \text{ and } c_{k_1} \neq c_{k_2},\\
\text{if } a_{j_1} = a_{j_2} & \text{then } s_{i_1} \neq s_{i_2} \text{ and } c_{k_1} \neq c_{k_2},\\
\end{array} \right.
\end{equation}
\end{definition}

\begin{definition}[Hard Access Network Slicing]\label{def:HardAccessNetworkSlicing}
	\emph{Hard network slicing of $A$} is a set of triplets $T_{hard, A} = \{(s_i, a_j, c_k) : s_i \in S, a_j \in A, c_k \in C\},$ such that for any two triplets  $(s_{i_1}, a_{j_1}, c_{k_1}), (s_{i_2}, a_{j_2}, c_{k_2}) \in T_{hard, A}$ it holds: 
	\begin{equation}\label{eq:HardAccessNetworkSlicing}
	\left\{
	\begin{array}{rl}
	\text{if } s_{i_1} = s_{i_2} & \text{then } a_{j_1} \neq a_{j_2} \text{ and } c_{k_1} \neq c_{k_2},\\
	\text{if } c_{k_1} = c_{k_2} & \text{then } s_{i_1} \neq s_{i_2} \text{ and } a_{j_1} \neq a_{j_2}.
	\end{array} \right.
	\end{equation}
\end{definition}

\begin{theorem}\label{Thm:CombinatorialDesignsHardCoreNetworkSlicing}
$T_{hard,C} = \{(s_i, a_j, c_k) : s_i \in S, a_j \in A, c_k \in C\}$ is a hard network slicing, if and only if there exist a partial $(S',A',C')-$conjugate Latin rectangle where $S' \subseteq S$, $A' \subseteq A$ and $C' \subseteq C$.
\end{theorem}
\textbf{Proof:}
%Let us first note that by convention the triplet of index sets $(E_1, E_3, E_3)$ in Definition \ref{def:conjugates} is the triplet $(E_1, E_3, E_3) \equiv (S, A, C)$.
If we are given a hard network slicing $T_{hard,C}$, then we can build an array $L$ as in Table \ref{table:UxA}, where the row indexing is by $s_i$ elements in $T_{hard,C}$ that forms a subset $S' \subseteq S$, column indexing is by $a_j$ elements in $T_{hard,C}$ that forms a subset $A' \subseteq A$, and entries by $c_k$ elements in $T_{hard,C}$ that form a subset $C' \subseteq C$. Due to Equation (\ref{eq:HardCoreNetworkSlicing}) in Definition \ref{def:HardCoreNetworkSlicing} it follows that the cells in $L$ are either empty or contain exactly one symbol, and no symbol occurs more than once in any row or column. Thus, the array obtained from $T_{hard}$ is a partial Latin rectangle.

Let $L$ be a partial $(S,A,C)-$conjugate Latin rectangle. Then we can build a set of triplets $T_{hard,C} = \{(s_i, a_j, c_k) : s_i \in S, a_j \in A, c_k \in C\},$ from the non-blank cells in $L$ such that Equation (\ref{eq:HardCoreNetworkSlicing}) holds.
$\blacksquare$

Definition \ref{def:HardCoreNetworkSlicing}, Definition \ref{def:HardAccessNetworkSlicing} and Theorem \ref{Thm:CombinatorialDesignsHardCoreNetworkSlicing} address the modeling of the hard core slicing with the $(S,A,C)$--conjugate. However, in practice we have network slices with components that are of mixed nature: sometimes a network slice has both core network and access network components as hard components, but sometimes one or both of those components are shared. That situation is best modeled with the $(C,A,S)$--conjugate rectangles, as shown in the next Theorem. 

\begin{theorem}\label{Thm:CAS-Reordering}
	Let all network slices are represented as a set of triplets $T = \{(c_i, a_j, s_k) : c_i \in C, a_j \in A, s_k \in S\}$, where $i\in\{1,\ldots,n_c\}$, $j\in\{1,\ldots,n_a\}$ and $k\in\{1,\ldots,n_s\}$. Then, there is a rectangular array $\mathcal{R}_{n_c \times n_a}$ of type $(C, A, S)$ and size $n_c \times n_a$ 
	%(i.e. array where rows are indexed by the elements of $C$, columns are indexed by the elements of $A$, and the cells contain none, one or several elements of $S$), 
	and there are values $1\le n_1 \le n_c$ and $1\le n_2 \le n_a$ such that the array is partitioned in four rectangular sub-arrays 
	\begin{equation}
		\mathcal{R}_{n_c \times n_a} = 
\begin{array}{|c||c|c|}
\hline
 & \multicolumn{2}{c|}{$A$} \\ \cline{2-3}
\hline
\hline
\multirow{2}{*}{$C$} & \mathcal{R}_{1,1} & \mathcal{R}_{1,2} \\ \cline{2-3}
%\hline 
\ & \mathcal{R}_{2,1} & \mathcal{R}_{2,2} \\ \cline{2-3}
\hline
\end{array}
	\end{equation}
	where $\mathcal{R}_{1,1} \equiv \mathcal{R}_{n_1 \times n_2}$, $\mathcal{R}_{1,2} \equiv \mathcal{R}_{n_1 \times (n_a - n_2)}$, $\mathcal{R}_{2,1} \equiv \mathcal{R}_{(n_c - n_1) \times n_2}$, $\mathcal{R}_{2,2} \equiv \mathcal{R}_{(n_c - n_1) \times (n_a - n_2)}$, and following holds:
	\begin{enumerate}
		\item every row and every column in $\mathcal{R}_{1,1}$ have at most one non-empty cell;
		\item every row in $\mathcal{R}_{1,2}$ has at most one non-empty cell, but its columns can have none, one or several non-empty cells;
		\item every column in $\mathcal{R}_{2,1}$ has at most one non-empty cell, but its rows can have none, one or several non-empty cells;
		\item every column and every row in $\mathcal{R}_{2,2}$ can have none, one or several non-empty cells.
	\end{enumerate}
\end{theorem}
\textbf{Proof:}
	Let us reorder the elements of $C$ as follows: $C_{hard} = \{ c_1, \ldots, c_{n_1} \}$ are components from the core network part that can be used only as dedicated, i.e., hard slicing, $C_{soft} = \{ c_{n_1+1}, \ldots, c_{n_c} \}$ are components that can be shared among NSIs. Then it is clear that $C = C_{hard} \cup C_{soft}$ is represented as a disjunctive union of dedicated and shared core network components. Let us apply the same reordering for the components in the access part, i.e., let us represent $A = A_{hard} \cup A_{soft}$ where $A_{hard} = \{ a_1, \ldots, a_{n_2} \}$ and $A_{soft} = \{ a_{n_2+1}, \ldots, a_{n_a} \}$. With this reordering for every slice $(c_i, a_j, s_k) \in T$ it holds:
	\begin{equation*}
	\left\{
	\begin{array}{rl}
	s_k \in \mathcal{R}_{1,1} & \text{if } 1\le i \le n_1 \text{\ and } 1\le j \le n_2,\\
	s_k \in \mathcal{R}_{1,2} & \text{if } 1\le i \le n_1 \text{\ and } n_2 + 1\le j \le n_a,\\
	s_k \in \mathcal{R}_{2,1} & \text{if } n_1 + 1\le i \le n_c \text{\ and } 1\le j \le n_2,\\
	s_k \in \mathcal{R}_{2,2} & \text{if } n_1 + 1\le i \le n_c \text{\ and } n_2 + 1\le j \le n_a.\\
	\end{array} \right.
	\end{equation*}
	Thus, for $s_k \in \mathcal{R}_{1,1}$ we can apply both conditions (\ref{eq:HardCoreNetworkSlicing}) and (\ref{eq:HardAccessNetworkSlicing}) from Definitions \ref{def:HardCoreNetworkSlicing} and \ref{def:HardAccessNetworkSlicing}, and claim 1 from Theorem \ref{Thm:CAS-Reordering} will follow. To see the validity of the claim 2 for $s_k \in \mathcal{R}_{1,2}$ we need only to apply the condition (\ref{eq:HardCoreNetworkSlicing}). Similarly, for the validity of the claim 3 and $s_k \in \mathcal{R}_{2,1}$ we need only to apply the condition (\ref{eq:HardAccessNetworkSlicing}). Then, the correctness of the remaining final claim 4 when $s_k \in \mathcal{R}_{2,2}$ follows.
$\blacksquare$

\begin{example}
	Let us represent network slicing case presented in Fig. \ref{architecture} and Table \ref{table:CxA} as a table following Theorem \ref{Thm:CAS-Reordering}.
\begin{table}[h!]
	\caption{A rectangular scheme equivalent to Table \ref{table:CxA}.}
	\centering
	\begin{tabular}{|l||c|c|c||c|c|c|}
		\hline
		                & $a_1$ & $a_2$ & $a_5$& $a_3$     & $a_4$ & $a_6$ \\
		\hline  \hline
		         $c_1$  & $s_1$ &       &      &           &       &       \\ \cline{1-7}
		         $c_2$  &       & $s_2$ &      &           &       &       \\ \cline{1-7}
		         $c_5$  &       &       & $s_4$&           &       &       \\ \cline{1-7}
		\hline
		\hline
		         $c_3$  &       &       &      & $s_2, s_3$ &       &       \\ \cline{1-7}
		         $c_4$  &       &       &      &           & $s_3$  & $s_5$ \\ \cline{1-7}
		         $c_6$  &       &       &      &           & $s_4$  &       \\ \cline{1-7}
		\hline
	\end{tabular}
	\label{table:CxA-TheoremReordering}
\end{table}
\end{example}

\begin{table*}[h!]
	\caption{A list of all components used by an NSMF for network slices given in a form of an extended $(C,A,S)$--conjugate as described by the attributes in expression (\ref{eq:NetworkSLice01})}
	\centering
	%\small
	\begin{tabular}{|l|c|c|c|} 
		\parbox{1.0cm}{Notation} & \parbox{3.8cm}{\vspace{0.1cm}\centering Meaning\vspace{0.1cm}} & \parbox{4.8cm}{\centering Relations/functions} & \parbox{5.5cm}{\centering Comment} \\
		\hline
		\hline
		$(c, a, s, t_{s}, t_{w})$ & \parbox{3.8cm}{Network slice}  & \parbox{4.8cm}{\vspace{0.1cm} $c \in C$, $a \in A$, $s\in S$, \\ $t_{s}$ - remaining life time of the slice, \\$t_{w}$ - waiting time before slice was activated. \vspace{0.1cm}} & \parbox{5.5cm}{\vspace{0.1cm} Initial value of $t_s$ is a random variable with exponential distribution and average value of $\mu$ time units. In the simulation, $t_s$ is decreased by 1 in every time unit.  \vspace{0.1cm}}\\
		\hline
		$\mu$ & \parbox{3.8cm}{An average life time of a network slice expressed in  number of time units and modeled with an exponential distribution}  & \parbox{4.8cm}{\vspace{0.1cm} $P(t_s \le x) = \left\{
			\begin{array}{rl}
			1-e^{-\frac{x}{\mu }} & \text{if } x \ge 0,\\
			0 & \text{if } x < 0.
			\end{array} \right.$ \\is the probability that the value of $t_s$ is less or equal to some value $x$ i.e. its cumulative distribution function. \vspace{0.1cm}} & \parbox{5.5cm}{\vspace{0.1cm} Expected value of $t_s$ is $E[t_s] = \mu$.\vspace{0.1cm}}\\
		\hline
		$C_{hard}$ & \parbox{3.8cm}{Set of hard slice core network components}    & \parbox{4.8cm}{\vspace{0.1cm} $C_{hard} = \{ c_1, \ldots, c_{n_1} \}$,\ \ \  $|C_{hard}| = n_1$  \vspace{0.1cm} } & \parbox{5.5cm}{\linespread{0.8}\vspace{0.1cm}An important parameter for the set $C_{hard}$ is the number of its elements $n_1$.\vspace{0.1cm}} \\
		\hline
		$C_{soft}$ & \parbox{3.8cm}{Set of shared core network components}    & \parbox{4.8cm}{\vspace{0.1cm} $C_{soft} = \{ c_{n_1+1}, \ldots, c_{n_c} \}$,\ \ \  $|C_{soft}| = n_c - n_1$  \vspace{0.1cm} } & \parbox{5.5cm}{\linespread{0.8}\vspace{0.1cm}If we denote the total number of core network components with $n_c$ then the number of elements in $C_{soft}$ is given as  $n_c - n_1$.\vspace{0.1cm}} \\
		\hline
		$C$ & \parbox{3.8cm}{\vspace{0.1cm}Set of all core network components \vspace{0.1cm}}    & \parbox{4.8cm}{\vspace{0.1cm} $C = C_{hard} \cup C_{soft}$,\ \ \  $|C| = n_c$  \vspace{0.1cm} } & \parbox{5.5cm}{\linespread{0.8}\vspace{0.1cm}Total number of core network components is $n_c$.\vspace{0.1cm}} \\
		\hline
		$A_{hard}$ & \parbox{3.8cm}{Set of hard slice access network components}    & \parbox{4.8cm}{\vspace{0.1cm} $A_{hard} = \{ a_1, \ldots, a_{n_2} \}$,\ \ \  $|A_{hard}| = n_2$  \vspace{0.1cm} } & \parbox{5.5cm}{\linespread{0.8}\vspace{0.1cm}An important parameter for the set $A_{hard}$ is the number of its elements $n_2$.\vspace{0.1cm}} \\
		\hline
		$A_{soft}$ & \parbox{3.8cm}{Set of shared access network components}    & \parbox{4.8cm}{\vspace{0.1cm} $A_{soft} = \{ a_{n_2+1}, \ldots, a_{n_a} \}$,\ \ \  $|A_{soft}| = n_a - n_2$  \vspace{0.1cm} } & \parbox{5.5cm}{\linespread{0.8}\vspace{0.1cm}If we denote the total number of access  network components with $n_a$ then the number of elements in $A_{soft}$ is given as  $n_a - n_2$.\vspace{0.1cm}} \\
		\hline
		$A$ & \parbox{3.8cm}{Set of all access network components}    & \parbox{4.8cm}{\vspace{0.1cm} $A = A_{hard} \cup A_{soft}$,\ \ \  $|A| = n_a$  \vspace{0.1cm} } & \parbox{5.5cm}{\linespread{0.8}\vspace{0.1cm}Total number of access  network components is $n_a$.\vspace{0.1cm}} \\
		\hline
		$S$ & \parbox{3.8cm}{\vspace{0.1cm}Set of all established network slices.\vspace{0.1cm}}    & \parbox{4.8cm}{\vspace{0.1cm} $S =  \{ s_1, \ldots, s_{n_s} \}$,\ \ \  $|S| = n_s$  \vspace{0.1cm} } & \parbox{5.5cm}{\linespread{0.8}\vspace{0.1cm}Number of active network slices in one particular moment $t$ is $n_s$. Notice, that in the next time moment $t + 1$ the number $n_s$ might change. \vspace{0.1cm}} \\
		\hline
		\parbox{1.0cm}{$req = (c, a, s, t_{s}, 1)$} & \parbox{3.8cm}{Initial request for a network slice}  & \parbox{4.8cm}{\vspace{0.1cm} $c \in C$, $a \in A$, $s\in S$ \vspace{0.1cm}} & \parbox{5.5cm}{\vspace{0.1cm} If NSMF decides that there are no resources for this request, $t_w$ is increased by 1, and the request is put back to the waiting queue for the next time unit.  \vspace{0.1cm}}\\
		\hline
		$p_c$ & \parbox{3.8cm}{\vspace{0.1cm}For a requested slice $req = (c, a, s, t_{s}, 0)$ the probability that $c \in C_{hard}$ \vspace{0.1cm}}    & \parbox{4.8cm}{\vspace{0.1cm} $P(c \in C_{hard}) = p_c$ \\
		$P(c \in C_{soft}) = 1 - p_c$  \vspace{0.1cm} } & \parbox{5.5cm}{\linespread{0.8}\vspace{0.1cm}Since dedicated resources are more expensive to use, the probability for requests that will ask for dedicated core network components is usually $p_c<0.5$.\vspace{0.1cm}} \\
		\hline
		$p_a$ & \parbox{3.8cm}{\vspace{0.1cm}For a requested slice $req = (c, a, s, t_{s}, 0)$ the probability that $a \in A_{hard}$ \vspace{0.1cm}}    & \parbox{4.8cm}{\vspace{0.1cm} $P(a \in A_{hard}) = p_a$ \\
		$P(a \in A_{soft}) = 1 - p_a$  \vspace{0.1cm} } & \parbox{5.5cm}{\linespread{0.8}\vspace{0.1cm}The probability for requests that will ask for dedicated access network components is usually $p_a<0.5$.\vspace{0.1cm}} \\
		\hline
		$N_{req}$ & \parbox{3.8cm}{\vspace{0.1cm}Number of received requests for network slice in certain time moment $t$ \vspace{0.1cm}}    & \parbox{4.8cm}{\vspace{0.1cm} $Req = \{ req_1, \ldots, req_{\tiny N_{req}} \}$,\ \ \  $|Req| = N_{req}$  \vspace{0.1cm} } & \parbox{5.5cm}{\linespread{0.8}\vspace{0.1cm} $N_{req}$ is a random variable with Poisson distribution and average value of $\lambda$.\vspace{0.1cm}} \\
		\hline
		$\lambda$ & \parbox{3.8cm}{\vspace{0.1cm}An average number of received requests for network slices in certain time moment $t$, modeled with a Poisson distribution \vspace{0.1cm}}  & \parbox{4.8cm}{\vspace{0.1cm} $P(N_{req} = k) = \frac{e^{-\lambda } \lambda ^k}{k!}$. \vspace{0.1cm}} & \parbox{5.5cm}{\vspace{0.1cm} Expected value of $N_{req}$ is $E[N_{req}] = \lambda$.\vspace{0.1cm}}\\
		\hline
		\hline
	\end{tabular}
	\label{table:all-components-for-NSMF}
\end{table*}

\begin{definition}\label{def:ExtendedConjugate}
	We say that a network slice is represented in \emph{extended $(C,A,S)$--conjugate form} if it is given as tuple $(c, a, s, attr_1, \ldots, attr_l)$ where $c \in C$, $a \in A$ and $s \in S$ and $attr_{\nu}$ are some additional attributes that are considered as important features of the slice.
\end{definition}

\section{Simulation of NSMF with several optimization objectives} \label{simul}

Equipped with Theorem \ref{Thm:CAS-Reordering} and Definition \ref{def:ExtendedConjugate} we can implement and simulate any realistic scenario for NSMF. 
We assume that requests for resources in the AN and CN parts for implementing slices with different requirements arrive according to Poisson distribution with arrival rate $\lambda$ in each time unit. %Here we consider a time-frame-based processing of tenant requests: in every time frame, a random number of requests for (resources in the AN and CN parts)/slices arrive by tenants. Once a service request arrives at CSMF, CSMF translates the requirements into network requirements and communicates them to NSMF. CSMF/NSMF decides based on the service requirements whether dedicated or shared network resources are needed. 
NSMF checks if the pool of resources can support the creation of the slice. If not, then the request is re-queued for the next time unit. Upon acceptance, the NSMF creates a new NSI and allocates a corresponding resource bundle (NSSI AN and NSSI CN) to the new NSI.
We consider dynamic deployment where slices have life time of $\nu$ time units distributed with exponential distribution, and the resources allocated to the slices will be released and added back to the resource pool when the slice is deactivated. 

By choosing different types of attributes we have opportunity to model different objectives (one or several) of the NSMF such as:
\begin{enumerate}
	\item to maximize the utilization of the network components;
	\item to decrease the average delay time from slice request to slice activation;
	\item to decrease the number of rejected slice requests;
	\item to maximize network operator revenue;
	\item to maximize the number of slices with high throughput.
\end{enumerate}  

In this section we give simulation results of an implementation of a NSMF for simple network slicing described with the following attributes:
\begin{eqnarray}\label{eq:NetworkSLice01}
\text{High level abstraction of a Network Slice Instance} \nonumber\\ 
(c, a, s, t_{s}, t_{w}) \hspace{1.6cm}
\end{eqnarray}
where $t_{s}$ is the remaining life time of the slice, $t_{w}$ is the time passed from the slice request until the slice was activated. By default $t_{w}=1$ when the request is composed. A full description of all components necessary for implementation of the NSFM is given in Table \ref{table:all-components-for-NSMF}.

\textbf{Note:} With the attribute list described in expression (\ref{eq:NetworkSLice01}) we work with a NSMF model where all hard resources and all soft resources from the core network and the access network are picked from a pool of resources. NSMF in this model has a higher level of abstraction and it does not take into account the specific capacity of the requested resources. Still, as we will show further in this work, even with this very abstracted model, we can infer important conclusions about the functionality of the network slicing concept and NSMF.
Nevertheless, our combinatorial model of network slicing can describe more detailed variants of NSMF. For example, 
\begin{eqnarray}\label{eq:NetworkSLiceMoreDetails}
\text{A Network Slice with quantitative resources} \nonumber\\ 
(c, a, s, t_{s}, t_{w}, r_{c}, r_{a}) \hspace{1.3cm}
\end{eqnarray}
where $t_{s}$ is the remaining life time of the slice, $t_{w}$ is the time passed from the slice request until the time the slice was activated, $r_{c}$ is the quantitative value requested from the core network and $r_{a}$ is the quantitative value requested from the access network.

We now give the algorithm that simulates the work of NSMF with network slices described with the expression (\ref{eq:NetworkSLice01}) and a scenario where rejected requests are added in the waiting queue to be considered for scheduling in the next time unit. Those rejected requests will compete for the network resources with the newly arrived requests.

\begin{algorithm}
	\caption{Simulation of NSMF with dynamic deployment and re-queuing of rejected requests.}\label{Alg:NSMF-Re-queuing}
	\begin{algorithmic}[1]
		\State $ActiveSlices \gets \emptyset$
		\State $RejReq \gets \emptyset$
		\State $n_s \gets 0$
		\For{$t = 1$ to $TimeSimulation$}
			\State $N_{req} \gets Poisson(\lambda)$
			\State $Req \gets \mathsf{GetRequests}[N_{req}, \mu, p_c, p_a] \cup RejReq$
			\State $RejReq \gets \emptyset$
			\State $Req \gets \mathsf{HeuristicRearangement}[Req]$
			\For{$req = (c, a, s, t_s, t_w) \in Req$ }
				\State $(Found_C, Found_A) \gets \mathsf{Dispetch}[req, C, A]$
				\If {$Found_C$ AND $Found_A$}
					\State $ActiveSlices \gets ActiveSlices \cup \{req\} $
					\State $C \gets C \setminus \{req.c\} $
					\State $A \gets A \setminus \{req.a\} $
				\Else
					\State $req.t_w \gets req.t_w + 1$
					\State $RejReq \gets RejReq \cup \{req\}$
				\EndIf
			\EndFor
			
			\State $NewActive \gets \emptyset$
			\For{$req = (c, a, s, t_s, t_w) \in ActiveSlices$ }
				\State $req.t_s \gets req.t_s - 1$
				\If {$req.t_s > 0 $}
					\State $NewActive \gets NewActive \cup \{req\} $
				\Else
					\State $C \gets C \cup \{req.c\} $
					\State $A \gets A \cup \{req.a\} $
				\EndIf				
			\EndFor
			\State $ActiveSlices \gets NewActive$
		\EndFor
	\end{algorithmic}
\end{algorithm}

In Algorithm \ref{Alg:NSMF-Re-queuing} we use several sub-functions that we comment here. In Step 5 the variable $N_{req}$ gets a random value according to a Poison distribution with a parameter $\lambda$. In Step 6 the function $\mathsf{GetRequests}[N_{req}, \mu, p_c, p_a]$ returns a set of initial requests $Req$, according to the parameters $N_{req}$, $\mu$, $p_c$, $p_a$ as they are described in Table \ref{table:all-components-for-NSMF}. 

In Step 8 there is a call to a procedure that rearranges the active list of requests $Req \gets \mathsf{HeuristicRearangement}[Req]$. That rearrangement can return just the original list of requests if we do not have developed any optimization strategy, or it can perform some heuristics in order to achieve better results with the next subroutine $\mathsf{Dispetch}[req, C, A]$ called in Step 10. Based on the rearrangement described in Theorem \ref{Thm:CAS-Reordering} we have developed one very simple but effective heuristics described in Algorithm \ref{Alg:HeuristicRearangement}. The idea can be briefly described as: give priorities to requests that belong to the rectangles $\mathcal{R}_{1,1}$, then $\mathcal{R}_{1,2}$ and $\mathcal{R}_{2,1}$ and finally to the rectangle $\mathcal{R}_{2,2}$. Within the subsets of requests in these rectangles, give priority to the requests that will finish sooner rather then later (that is sorting in ascending order in Steps 16 - 19).

\begin{algorithm}
	\caption{$\mathsf{HeuristicRearangement}[Req]$}\label{Alg:HeuristicRearangement}
	\begin{algorithmic}[1]
		\State $Req_{1,1} \gets \emptyset$, $Req_{1,2} \gets \emptyset$, $Req_{2,1} \gets \emptyset$, $Req_{2,2} \gets \emptyset$
		\For{$req = (c, a, s, t_s, t_w) \in Req$ }
			\If {$c\in C_{hard}$ AND $a \in A_{hard}$}
				\State $Req_{1,1} \gets Req_{1,1} \cup req$
			\EndIf
			\If {$c\in C_{hard}$ AND $a \in A_{soft}$}
				\State $Req_{1,2} \gets Req_{1,2} \cup req$
			\EndIf
			\If {$c\in C_{soft}$ AND $a \in A_{hard}$}
				\State $Req_{2,1} \gets Req_{2,1} \cup req$
			\EndIf
			\If {$c\in C_{soft}$ AND $a \in A_{soft}$}
				\State $Req_{2,2} \gets Req_{2,2} \cup req$
			\EndIf
		\EndFor
		\State $Req_{1,1} \gets \mathsf{Sort Ascending}[Req_{1,1}, t_s]$		
		\State $Req_{1,2} \gets \mathsf{Sort Ascending}[Req_{1,2}, t_s]$		
		\State $Req_{2,1} \gets \mathsf{Sort Ascending}[Req_{2,1}, t_s]$		
		\State $Req_{2,2} \gets \mathsf{Sort Ascending}[Req_{2,2}, t_s]$
		\State $Req \gets Req_{1,1} || Req_{1,2} || Req_{2,1} || Req_{2,2}$
		\State \textbf{Return} $Req$
	\end{algorithmic}
\end{algorithm}

If $\mathsf{Dispetch}[ ]$ subroutine returns that there are resources both in the core network and in the access network, then that request is activated by adding it in Step 12 to the list of active slices, and the list of core network and access network resources is updated in Steps 13 and 14. If $\mathsf{Dispetch}[ ]$ subroutine returns that there are no available resources then the waiting time for the request is increased by one, and the rejected request is added to the set of rejected requests.

Steps 20 to 30 update the state of the active slices by reducing by 1 all their $t_s$ values. If the slice has a value $t_s$ that is still positive, it will continue to be active for the next time unit. Otherwise, the slice is deactivated and its resources are released and are added back in the pool of available resources (Steps 26 and 27). 

\begin{lemma}\label{Lemma:StabilityConditions}
	Necessary conditions for Algorithm \ref{Alg:NSMF-Re-queuing} to reach a stationary ergodic stage are the following: 
	\begin{align}\label{eq:p_c}
	p_c & < \frac{n_1}{n_c},\\
	p_a & < \frac{n_2}{n_a}, \\
	\mu  \lambda & < \min\{n_c, n_a\}.
	\end{align}
\end{lemma}
\textbf{Proof:}
	(sketch) The proof is by assuming that any of the given inequalities is not true and by showing in that case Algorithm \ref{Alg:NSMF-Re-queuing} produces an ever increasing list of requests $Req$. For example, let us assume that $p_c \ge \frac{n_1}{n_c}$. This means that in average, there will be more requests asking for hard core network components than there are available, thus rejecting those requests, i.e., producing longer and longer requests lists $Req$. 
	
	A similar reasoning is if we suppose that $\mu  \lambda \ge \min\{n_c, n_a\}$. This means that there will be a situation when the number of requests times the number of time units necessary to finish the activity of those requests will surpass the minimum number of available resources either in the core part or in the access part. In that case the rejected requests will be added to the queue of requests, thus contributing for ever-increasing length of the list of requests $Req$.
$\blacksquare$

We have an initial implementation of Algorithms \ref{Alg:NSMF-Re-queuing} and \ref{Alg:HeuristicRearangement} in Mathematica, and next we show several experimental results that confirm the claims of Lemma \ref{Lemma:StabilityConditions}, especially the effects of compliance vs non-compliance with the conditions (\ref{eq:p_c}), (7) and (8).

\begin{figure}
	%\centering
	\includegraphics[width=3.3in]{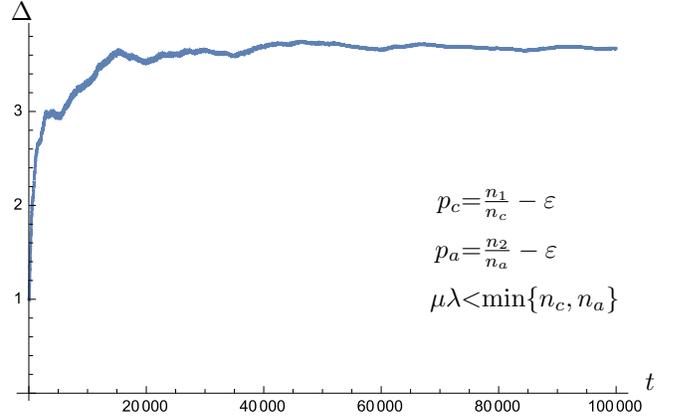}
	\put(-90,60){
	$\begin{array}{r@{}c@{}l}
	   p_c  & =  & \frac{n_1}{n_c} - \varepsilon \vspace{0.2cm} \\
	   p_a  & =  & \frac{n_2}{n_a}  - \varepsilon \vspace{0.2cm}\\
	   \mu  \lambda & < & \min\{n_c, n_a\}
	\end{array}
	$}
	\put(0,10){$t$}
	\put(-240,150){$\Delta$}
	\caption{An average activation delay simulating the work of NSMF for 100,000 time units. The average is taken over 10 experiments. After a transitioning phase of about 15,000 time units, the process becomes stationary ergodic and the average delay $\Delta$ is around 3.5 }
	\label{Fig:ActivationDelay-Compliance099}
\end{figure}

\begin{figure}
	\centering
	\includegraphics[width=3.3in]{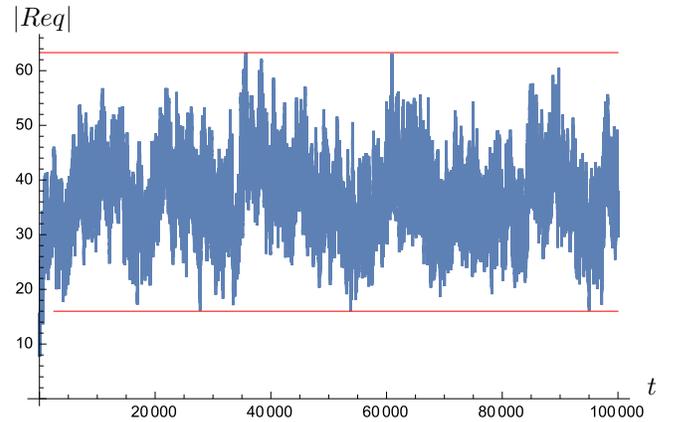}
	\put(0,10){$t$}
	\put(-240,150){$|Req|$}
	\caption{An average request queue size simulating the work of NSMF for 100,000 time units. The average is taken over 10 experiments. The size of the requests  queue $|Req|$ is stationary ergodic and varies between 16 and 63.}
	\label{Fig:ReqQueueSizeComplieance099}
\end{figure}

In Fig. \ref{Fig:ActivationDelay-Compliance099} and Fig. \ref{Fig:ReqQueueSizeComplieance099} we give the results of performing 10 simulations with the following parameters: $n_1 = 50$, $n_c = 350$, $p_c = 0.99 \frac{n_1}{n_c} = 0.141429$, $n_2 = 100$, $n_a = 500$, $p_a = 0.99 \frac{n_2}{n_a} = 0.198$, $\lambda = 10$, $\mu = \lfloor 0.99 \frac{\min\{n_c, n_a\}}{\mu} \rfloor = 34$. The simulation was performed for 100,000 time units. As we can see in Fig. \ref{Fig:ActivationDelay-Compliance099}, there is a transition period of about 15,000 time units until the process becomes stationary ergodic with an average delay $\Delta$ around 3.5 time units. In Fig. \ref{Fig:ReqQueueSizeComplieance099} we show the corresponding queue size for the same simulation. The size of the  queue $|Req|$ is stationary ergodic and varies between 16 and 63 requests. 

\begin{figure}
	\centering
	\includegraphics[width=3.3in]{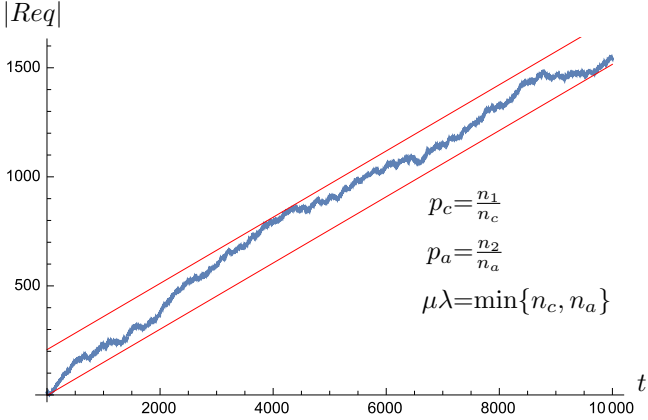}
	\put(-90,60){
	$\begin{array}{r@{}c@{}l}
	   p_c  & =  & \frac{n_1}{n_c} \vspace{0.2cm} \\
	   p_a  & =  & \frac{n_2}{n_a} \vspace{0.2cm}\\
	   \mu  \lambda & = & \min\{n_c, n_a\}
	\end{array}
	$}
	\put(0,10){$t$}
	\put(-240,150){$|Req|$}
	\caption{An average request queue size simulating the work of NSMF for 10,000 time units. The average is taken over 10 experiments. By having parameters that are upper bounds in Lemma \ref{Lemma:StabilityConditions}, the functioning of the NSMF is not a stable process since the size of the requests queue $|Req|$ is always increasing.}
	\label{Fig:ReqQueueSizeNonComplieance}
\end{figure}

In Fig. \ref{Fig:ReqQueueSizeNonComplieance} we show the results of 10 simulations with the values that are the upper bounds in Lemma \ref{Lemma:StabilityConditions}, i.e., $p_c  =  \frac{n_1}{n_c} = 0.2$, $p_a   =  \frac{n_2}{n_a} = 0.142857$ and  $\mu  \lambda  =  \min\{n_c, n_a\} = 35$. As we can see the size of the requests queue is always increasing as times goes on, indicating that the parameters chosen by the network operator are not sustainable in this model. This simulation analysis indicates also that the network operator should either increase the pool size for the access and core network resources in order to avoid the strict equality or some rejection policy should be introduced.

\shorten{Acceptance ratio is the ratio of the number of network slice requests which have been successfully mapped and the total number of network slice requests.}

\shorten{Imame li static (slices are permanent once they are deployed successfully) ili dynamic (slices have life time and the resources allocated to the slices will be recycled at the end of the life time) deployment }

As mentioned before, we can seek for several optimization objectives within one model of NSMF. Here we give results from simulation of optimized and non-optimized NSMF given by the NSI expression (\ref{eq:NetworkSLice01}), where the optimization is performed in Step 8 of the Algorithm \ref{Alg:NSMF-Re-queuing}, and where the optimization heuristics is given in Algorithm \ref{Alg:HeuristicRearangement}. The optimization objectives are: 1. to maximize the utilization of the network components; and 2. to decrease the average delay time from slice request to slice activation. We argue that by these objectives, indirectly we are achieving also the objective to maximize the network operator revenue.

\begin{definition}\label{Def:Utilization}
Let $U(t), t=1, \ldots$, be a function that denotes the number of network slice resources scheduled by the NSMF at time $t$. An average utilization $V_{[T_1,T_2]}$ of the network components for the NSMF model given by the NSI expression (\ref{eq:NetworkSLice01}), for the time period $[T_1, T_2]$ is defined as  
\begin{equation}\label{eq:Utilization}
    V_{[T_1,T_2]} = \frac{1}{T_2 - T_1}\sum_{t=T_1}^{T_2} U(t)
\end{equation}
\end{definition}

Without a proof we state here the following Corollary.
\begin{corollary}\label{col:UpperBoundOfUtilization}
For NSMF model given by the NSI expression (\ref{eq:NetworkSLice01}), for any time interval $[T_1,T_2]$, $V_{[T_1,T_2]}$ is upper bounded by $\min (m_c, n_a)$, i.e.,
\begin{equation}\label{eq:UpperBoundOnUtilization}
    V_{[T_1,T_2]} \le \min (m_c, n_a).
\end{equation} $\blacksquare$
\end{corollary}

Seeking for optimization strategies that will increase the average utilization of the network components is a desired goal, but is not the most rational optimization objective because it excludes the delay time between the request and the service delivery. Thus, it is much better to set another optimization objective which we define with the next definition.

\begin{definition}\label{def:UtilizationRatio}
Let $U(t), t=1, \ldots$, be a function that denotes the number of network slice resources scheduled by the NSMF at time $t$, and let $\Delta(t), t=1, \ldots$, be a function that denotes the average delay units that network slice requests issued at time $t$ should wait until their activation. 
An utilization ratio function $W(t)$ is defined as:
\begin{equation}\label{eq:UtilizationRatioFunction}
    W(t) = \frac{U(t)}{\Delta(t)} 
\end{equation}
An average utilization ratio $W_{[T_1,T_2]}$ for the time period $[T_1, T_2]$ is defined as  
\begin{equation}\label{eq:UtilizationRatio}
    W_{[T_1,T_2]} = \frac{1}{T_2 - T_1}\sum_{t=T_1}^{T_2} W(t)
\end{equation}
\end{definition}

Our objective is to define optimization strategies that maximize $W_{[T_1,T_2]}$ for any time interval $[T_1,T_2]$.

\begin{figure}
	\centering
	\includegraphics[width=3.3in]{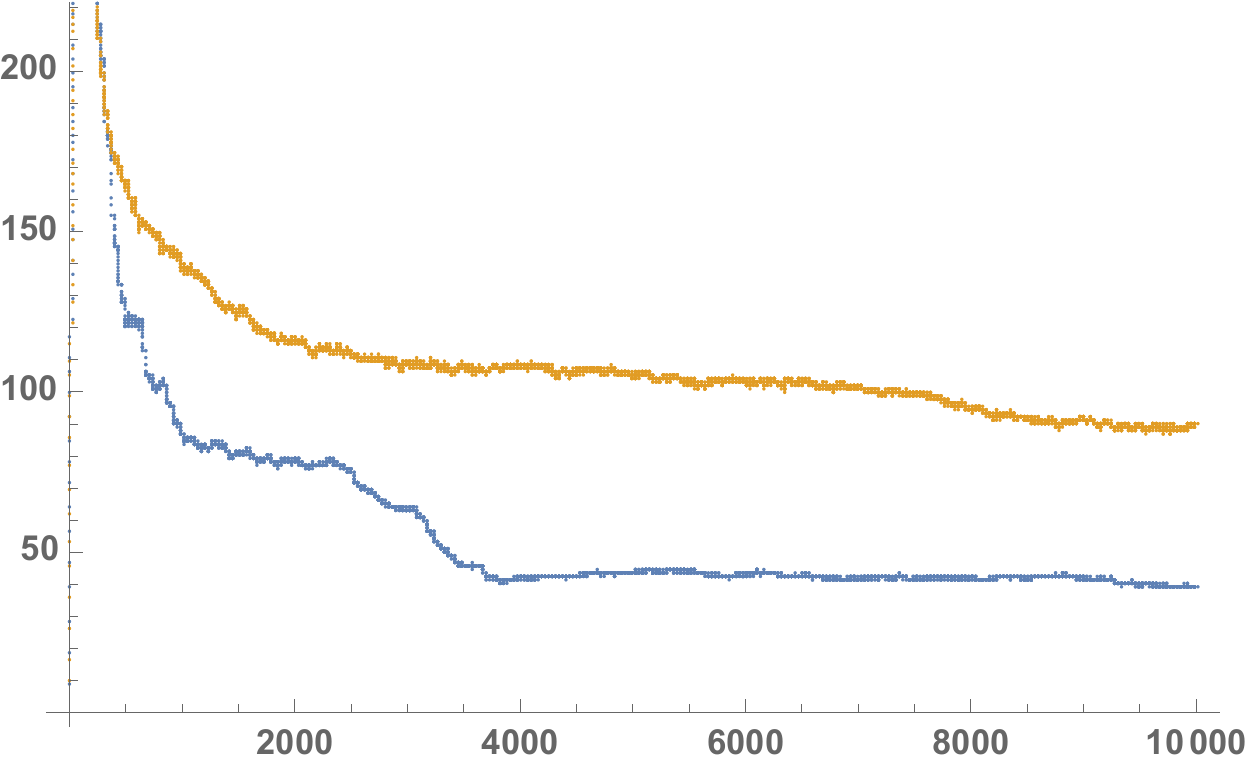}
	\put(-70,120){
	$\begin{array}{r@{}c@{}l}
	   p_c  & =  & 0.95 \frac{n_1}{n_c} \vspace{0.2cm} \\
	   p_a  & =  & 0.95 \frac{n_2}{n_a} \vspace{0.2cm}\\
	   \mu  & = & 10 \vspace{0.2cm}\\ 
	   \lambda & = & 34
	\end{array}
	$}
	\put(0,10){$t$}
	\put(-240,150){$W(t)$}
	\caption{Utilization ratio function simulating the work of NSMF for 10,000 time units. The average is taken over 10 experiments. The orange curve is obtained by the optimal strategy in Algorithm \ref{Alg:HeuristicRearangement} and the blue curve is for simulation without any optimizations (requests are processed as they arrive).}
	\label{Fig:UtilizationRatioComparison01}
\end{figure}

In Fig. \ref{Fig:UtilizationRatioComparison01} we show comparison between two utilization ratio functions where one is obtained without any optimization heuristics, i.e., the requests are processed as they come in a first-come-first-serve manner (the blue curve), and the other is obtained by the Algorithm \ref{Alg:HeuristicRearangement} (the orange curve). For the non-optimized version we get $W_{[4000, 10000]} = 42.248$ that means in every moment the ratio between the number of used resources and the waiting time is 42.248. On the other hand, the optimal strategy gives us $W_{[4000, 10000]} = 98.865$ which is more than double than the non-optimized strategy.

\section{Conclusion} \label{conc}
We proposed a mathematical model for network slicing based on combinatorial designs such as Latin squares and rectangles and their conjugate forms. These combinatorial designs allow us to model both soft and hard slicing in the access and core parts. Moreover, by the introduction of the extended attribute description our model can offer different levels of abstractions for NSMF that combines cross-domain NSSIs in one end-to-end NSI. 

From the optimization point of view, in this work we also introduced the notion of utilization ratio function, with aims to describe the functional dependencies between the number of used network resources and the waiting time for establishing the network slice. Then, we presented two strategies for the work of NSMF, a non-optimized first-come-first-serve strategy and an optimal strategy, where the objectives of the optimization are: 1. to maximize the utilization of the network components; and 2. to decrease the average delay time from slice request to slice activation. Simulations results presented in this work show that optimal strategy achieved by maximizing the utilization ratio function provides more than twice better performances in terms of the both objectives.

%\section{Conclusion}
%A conclusion section is not required. Although a conclusion may review the 
%main points of the paper, do not replicate the abstract as the conclusion. A 
%conclusion might elaborate on the importance of the work or suggest 
%applications and extensions. 

\bibliography{refer}

\begin{thebibliography}{10}
\providecommand{\url}[1]{#1}
\csname url@samestyle\endcsname
\providecommand{\newblock}{\relax}
\providecommand{\bibinfo}[2]{#2}
\providecommand{\BIBentrySTDinterwordspacing}{\spaceskip=0pt\relax}
\providecommand{\BIBentryALTinterwordstretchfactor}{4}
\providecommand{\BIBentryALTinterwordspacing}{\spaceskip=\fontdimen2\font plus
\BIBentryALTinterwordstretchfactor\fontdimen3\font minus
  \fontdimen4\font\relax}
\providecommand{\BIBforeignlanguage}[2]{{%
\expandafter\ifx\csname l@#1\endcsname\relax
\typeout{** WARNING: IEEEtran.bst: No hyphenation pattern has been}%
\typeout{** loaded for the language `#1'. Using the pattern for}%
\typeout{** the default language instead.}%
\else
\language=\csname l@#1\endcsname
\fi
#2}}
\providecommand{\BIBdecl}{\relax}
\BIBdecl

\bibitem{kim2018network}
D.~Kim and S.~Kim, ``{Network slicing as enablers for 5G services: state of the
  art and challenges for mobile industry},'' \emph{Telecommunication Systems},
  pp. 1--11, 2018.

\bibitem{EricssonAndBT-17}
E.~study, ``{An economic study of 5G network slicing for IoT service
  deployment},'' {Ericsson}, Tech. Rep., 2017.

\bibitem{alliance20155g}
{NGMN Alliance}, ``{5G white paper},'' \emph{Next generation mobile networks,
  white paper}, pp. 1--125, 2015.

\bibitem{3gpp2015-22-891}
3GPP, ``{Study on new services and markets technology enablers},'' \emph{3GPP
  TS 22.891 version 1.0.0}, September 2015.

\bibitem{3gpp}
------, ``{System Architecture for the 5G System},'' \emph{3GPP TS 23.501
  version 15.2.0 Release 15}, 2018.

\bibitem{3gpp2}
------, ``{5G; Management and orchestration; Concepts, use cases and
  requirements},'' \emph{3GPP TS 28.530 version 15.0.0 Release 15}, 10-2018.

\bibitem{DBLP:journals/corr/LiWPU16}
\BIBentryALTinterwordspacing
Q.~Li, G.~Wu, A.~Papathanassiou, and M.~Udayan, ``{An end-to-end network
  slicing framework for 5G wireless communication systems},'' \emph{CoRR}, vol.
  abs/1608.00572, 2016. [Online]. Available:
  \url{http://arxiv.org/abs/1608.00572}
\BIBentrySTDinterwordspacing

\bibitem{8334921}
A.~Kaloxylos, ``{A Survey and an Analysis of Network Slicing in 5G Networks},''
  \emph{IEEE Communications Standards Magazine}, vol.~2, no.~1, pp. 60--65,
  MARCH 2018.

\bibitem{ABIResearchAndIntel2018}
A.~white paper, ``{The Evolution of Network Slicing},'' {ABI Research and
  Intel}, Tech. Rep., November 2017.

\bibitem{geng2017network}
L.~Geng, J.~Dong, S.~Bryant, K.~Makhijani, A.~Galis, X.~de~Foy, and
  S.~Kuklinsk, ``{Network slicing architecture},'' \emph{Internet Engineering
  Task Force, Internet-Draft draft-geng-netslices-architecture-02}, 2017.

\bibitem{3gpp1}
3GPP, ``{Study on management and orchestration of network slicing for next
  generation network},'' \emph{3GPP TR 28.801 version 15.1.0 Release 15}, 2018.

\bibitem{regalado2011coined}
A.~Regalado, ``{Who coined ‘cloud computing’},'' \emph{Technology Review},
  vol.~31, 2011.

\bibitem{yang2004forwarding}
L.~Yang, R.~Dantu, T.~Anderson, and R.~Gopal, ``{Forwarding and control element
  separation (ForCES) framework},'' Tech. Rep., 2004.

\bibitem{etsi2013001}
G.~ETSI, ``{001:" Network Functions Virtualisation (NFV)},''
  \emph{Architectural framework}, 2013.

\bibitem{ghodsi2011information}
A.~Ghodsi, S.~Shenker, T.~Koponen, A.~Singla, B.~Raghavan, and J.~Wilcox,
  ``{Information-centric networking: seeing the forest for the trees},'' in
  \emph{Proceedings of the 10th ACM Workshop on Hot Topics in Networks}.\hskip
  1em plus 0.5em minus 0.4em\relax ACM, 2011, p.~1.

\bibitem{8329496}
W.~Guan, X.~Wen, L.~Wang, Z.~Lu, and Y.~Shen, ``{A Service-Oriented Deployment
  Policy of End-to-End Network Slicing Based on Complex Network Theory},''
  \emph{IEEE Access}, vol.~6, pp. 19\,691--19\,701, 2018.

\bibitem{8382171}
B.~{Han}, J.~{Lianghai}, and H.~D. {Schotten}, ``{Slice as an Evolutionary
  Service: Genetic Optimization for Inter-Slice Resource Management in 5G
  Networks},'' \emph{IEEE Access}, vol.~6, pp. 33\,137--33\,147, 2018.

\bibitem{8057045}
D.~{Bega}, M.~{Gramaglia}, A.~{Banchs}, V.~{Sciancalepore}, K.~{Samdanis}, and
  X.~{Costa-Perez}, ``{Optimising 5G infrastructure markets: The business of
  network slicing},'' in \emph{IEEE INFOCOM 2017 - IEEE Conference on Computer
  Communications}, May 2017, pp. 1--9.

\bibitem{Colbourn:2006:HCD:1202540}
C.~J. Colbourn and J.~H. Dinitz, \emph{{Handbook of Combinatorial Designs,
  Second Edition (Discrete Mathematics and Its Applications)}}.\hskip 1em plus
  0.5em minus 0.4em\relax Chapman \& Hall/CRC, 2006.

\bibitem{10.1007/978-3-642-40552-5_15}
K.~Kralevska, D.~Gligoroski, and H.~{\O}verby, ``Balanced xor-ed coding,'' in
  \emph{Advances in Communication Networking}, T.~Bauschert, Ed.\hskip 1em plus
  0.5em minus 0.4em\relax Berlin, Heidelberg: Springer Berlin Heidelberg, 2013,
  pp. 161--172.

\bibitem{6875415}
D.~{Gligoroski} and K.~{Kralevska}, ``Families of optimal binary non-mds
  erasure codes,'' in \emph{2014 IEEE International Symposium on Information
  Theory}, June 2014, pp. 3150--3154.

\bibitem{7034478}
K.~{Kralevska}, H.~{{\O}verby}, and D.~{Gligoroski}, ``Joint balanced source
  and network coding,'' in \emph{2014 22nd Telecommunications Forum Telfor
  (TELFOR)}, Nov 2014, pp. 589--592.

\bibitem{Colbourn1999ApplicationsOC}
C.~J. Colbourn, J.~H. Dinitz, and D.~R. Stinson, ``{Applications of
  Combinatorial Designs to Communications, Cryptography, and Networking},''
  \emph{London Mathematical Society Lecture Note Series}, vol. 187, 1999.

\end{thebibliography}
\bibliographystyle{IEEEtran}

\end{document}